\documentclass[5p]{elsarticle}
\usepackage{lineno}
\usepackage[breaklinks=true]{hyperref}
\hypersetup{colorlinks,linkcolor=red,citecolor=blue,urlcolor=blue}

\usepackage{natbib}
\bibpunct{(}{)}{;}{a}{}{,} 

\journal{New Astronomy}

\newcommand{\apjs}{ApJS}
\newcommand{\apj}{ApJ}
\newcommand{\aj}{AJ}

\newcommand{\aap}{A\&A}
\newcommand{\mnras}{MNRAS}

\newcommand{\pasp}{PASP}

\def\astrobj#1{#1}

\begin{document}

\begin{frontmatter}

\title{NSV 1907 –- a New Eclipsing, Nova-Like Cataclysmic Variable}


\author[label1,label2]{Stefan H{\"u}mmerich}
\author[label1]{Rainer Gr{\"o}bel}
\author[label1,label2,label3]{Franz-Josef Hambsch}
\author[label3,label4]{Franky Dubois}
\author[label5]{Richard Ashley}
\author[label5]{Boris T. G{\"a}nsicke}
\author[label3,label4,label6]{Siegfried Vanaverbeke}
\author[label1,label2]{Klaus Bernhard}
\author[label3]{Patrick Wils}

\address[label1]{Bundesdeutsche Arbeitsgemeinschaft f{\"u}r Ver{\"a}nderliche Sterne e.V. (BAV), Berlin, Germany}
\address[label2]{American Association of Variable Star Observers (AAVSO), Cambridge, USA}
\address[label3]{Vereniging Voor Sterrenkunde (VVS), Brugge, Belgium}
\address[label4]{Astrolab IRIS, Ieper, Belgium}
\address[label5]{Department of Physics, University of Warwick, Coventry, UK}
\address[label6]{Center for Plasma Astrophysics, University of Leuven, Belgium}

\begin{abstract}
\astrobj{NSV 1907}, formerly listed as an irregular variable in variability catalogues, was classified as an Algol-type eclipsing binary in the Catalina Surveys Periodic Variable Star Catalogue. We have identified \astrobj{NSV 1907} as an ultraviolet (UV) bright source using measurements from the GALEX space telescope and detected obvious out-of-eclipse variability in archival photometric data from the Catalina Sky Survey, which instigated a closer examination of the object. A spectrum and extensive multicolour photometric observations were acquired, from which we deduce that \astrobj{NSV 1907} is a deeply eclipsing, nova-like cataclysmic variable. Apart from the orbital variations (deep eclipses with a period of $P \approx 6.63$ hours), changes in mean brightness and irregular short-term variability (flickering) were observed. The presence of a secondary minimum at phase $\varphi \approx 0.5$ was established, which indicates a significant contribution of the companion star to the optical flux of the system. We find possible evidence for sinusoidal variations with a period of $P \approx 4.2$ d, which we interpret as the nodal precession period of the accretion disc. No outbursts or \astrobj{VY Scl}-like drops in brightness were detected either by the CSS or during our photometric monitoring. Because of its spectral characteristics and the observed variability pattern, we propose \astrobj{NSV 1907} as a new moderately bright long-period \astrobj{SW Sextantis} star. Further photometric and spectroscopic observations are encouraged.
\end{abstract}

\begin{keyword}
binaries: eclipsing \sep stars: variables: cataclysmic variables \sep stars: variables: SW Sex stars \sep stars: individual: NSV 1907
\end{keyword}

\end{frontmatter}

\section{Introduction}

Cataclysmic variables (CVs) are interacting binary systems which comprise a low-mass secondary star (mostly a red dwarf) losing material to a white dwarf (WD) primary star. They exhibit complex photometric variability, which is characterised by a variety of phenomena like e.g. eclipses, rapid oscillations, ellipsoidal modulation and abrupt and conspicuous brightenings (i.e. dwarf nova eruptions or nova outbursts). Other tell-tale signs of CVs are the presence of strong emission lines in their optical spectra, their blue colour and, especially in the case of magnetic CVs, their X-ray luminosity, all of which are made good use of in searches for this kind of variable stars. For a general review of CVs, the reader is referred to \citet{Warn95} and \citet{Hell01}.

Nova-like cataclysmic variables are characterised by high mass transfer rates and prominent steady state accretion discs and do not show large amplitude outbursts. Among this subgroup of CVs, the \astrobj{SW Sextantis} stars are set apart by several common traits, which include unusual V-shaped eclipse profiles, single-peaked emission lines exhibiting central
absorption dips around orbital phases $\varphi \approx 0.4–-0.7$, a substantial orbital phase lag (0.1--0.2 cycle) of the radial velocities of the Balmer lines and high-velocity emission S-waves with maximum blueshift near phase $\varphi \approx 0.5$ (\citealp{Thrs91,Rodr07a}). \astrobj{SW Sex} systems are not rare but dominate in the period range $2.8 \le P\mathrm{orb} \le 4$ hours (\citealp{Rodr07b,Schm12}), just above the period gap. Recently, evidence has been mounting that the \astrobj{SW Sex} phenomenon is an evolutionary stage in the life of CVs \citep{Schm12}.

Optical transient surveys such as the Catalina Real-time Transient Survey \citep[CRTS;][]{Drak09} and the All-Sky Automated Survey for SuperNovae \citep[ASAS-SN1;][]{Shap14} are routinely discovering new dwarf novae by the hundreds through detection of their outbursts. Equally, X-ray observatories like e.g. ROSAT \citep{Voge91}, INTEGRAL \citep{Uber03}  or Swift \citep{Cusu10} have contributed to the discovery of many magnetic CVs. 

However, our knowledge of the intrinsic population of the nova-like CVs, which neither show large amplitude outbursts nor the X-ray emission typical of their magnetic brethren, is probably still rather incomplete. Here we outline a new method of identifying nova-like variables, and report the identification of \astrobj{NSV 1907} as a candidate moderately bright long-period \astrobj{SW Sex} star. The method of identification and observations of our target star are presented in Section \ref{observations}. Our data analysis is described in Section \ref{data_analysis} and we conclude in Section \ref{conclusion}.

\section{Method of Identification and Observations} \label{observations}

\subsection{Method of Identification} \label{methods}
As has been pointed out above, nova-like cataclysmic variables are difficult to identify because of the absence of large-scale outbursts that are readily detected by optical transient surveys. However, the optical spectra of nova-like stars, including the \astrobj{SW Sex} stars, are characterised by very blue continua. It is therefore not surprising that a substantial fraction of this class of variables has been discovered in ultraviolet (UV) excess surveys \citep{Rodr07b}.

\begin{figure}[h!]
	\centering
		\includegraphics[width=0.5\textwidth]{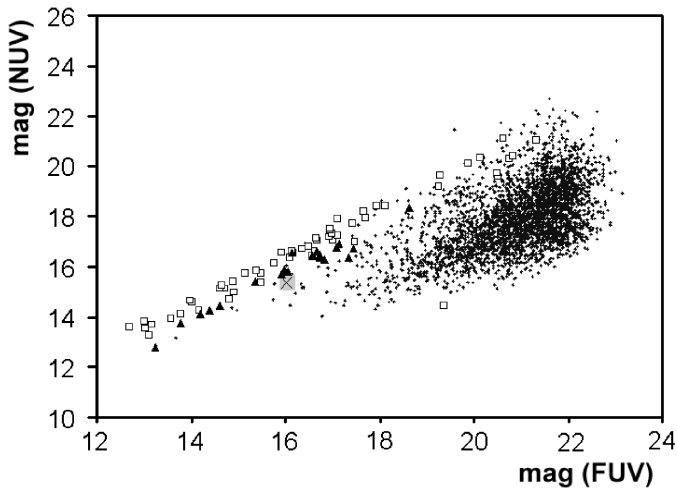}
	\caption{$FUV$ vs. $NUV$ diagram of all sources in the Catalina Surveys Periodic Variable Star Catalogue \citep{Drak14} with good GALEX photometry. Post-common envelope binaries are denoted by open squares. Definite or probable \astrobj{SW Sextantis} stars from the 'Big List of SW Sextantis Stars' \citep{Hoar03} are represented by filled triangles. The grey square marked with a cross indicates the position of \astrobj{NSV 1907}.}
	\label{figure1}
\end{figure}

As a starting point in the search for nova-like CVs, we have investigated the extensive sample of variable stars compiled in the Catalina Surveys Periodic Variable Star Catalogue \citep{Drak14} using UV photometry from the GALEX (Galaxy Evolution Explorer Space Telescope; \citealt{Mart05}) satellite, which has been monitoring the sky in $FUV$
($1344–-1786$ \AA, $\lambda\mathrm{eff} = 1538.6$ \AA) and $NUV$ ($1771-–2831$ \AA, $\lambda\mathrm{eff} = 2315.7$ \AA) simultaneously \citep{Morr07}. To this end, both source catalogues were cross-matched and the resultant list of objects was investigated in $FUV$ vs. $NUV$ colour space. The result is shown in Fig. \ref{figure1}.

Groups of interest were identified, like e.g. post--common envelope binaries (class code '16' in \citealt{Drak14}; denoted by open squares in Fig. \ref{figure1}) and definite or probable \astrobj{SW Sex} stars from the 'Big List of SW Sextantis Stars' \citep{Hoar03}\footnote{http://www.dwhoard.com/biglist} (represented by filled triangles in Fig. \ref{figure1}). UV bright objects situated near the expected loci of nova-like CVs were investigated in more detail. We concentrated in particular on stars that had been classified as short-period eclipsing binaries in the Catalina Surveys Periodic Variable Star Catalogue, which brought to our attention \astrobj{NSV 1907} ($FUV = 16.028$ mag; $NUV = 16.043$ mag; $(FUV-NUV) \approx 0$; position indicated by the grey square marked with a cross in Fig. \ref{figure1}). Inspection of the corresponding light curve from the Catalina Sky Survey (CSS; \citealt{Drak09}) resulted in the detection of obvious out-of-eclipse variability, which instigated a closer examination of the object.

\begin{figure}[h]
	\centering
		\includegraphics[width=0.5\textwidth]{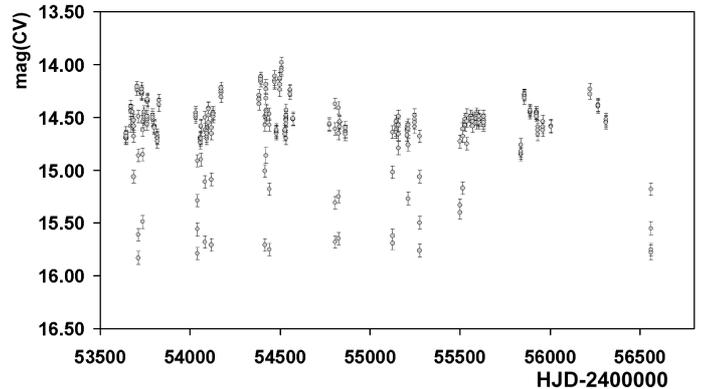}
	\caption{Light curve of \astrobj{NSV 1907}, based on data from the Catalina Sky Survey \citep{Drak09}.}
	\label{Fig0}
\end{figure}

\subsection{Target Star} \label{target_star}
The variability of GSC\,00104-02193 = USNO-B1.0\,0935-0067535 = 2MASS\,J05165408+0332525 (RA,\,Dec\,(J2000) = 05h\,16m\,54s.085, +03$^\circ$\,32'\,52''.45; UCAC4 position) was discovered by C. Hoffmeister on Sonneberg plates and the star was announced as a long-period variable with an amplitude of 0.5 mag \citep{Hoff63}. The star was primarily designated as S 8003 and later included in the New Catalogue of Suspected Variable Stars \citep{KuKh82} as \astrobj{NSV 1907}. No more information on the type of variability had been available until the object was included in the Catalina Surveys Periodic Variable Star Catalogue \citep{Drak14} under the designation of CSS J051654.1+033252 and classified as an Algol-type eclip\-sing binary (GCVS-type EA) with a period of $P = 0.2761060$ d and an eclipse depth of 1.10 mag.

\subsection{Archival Photometry} \label{arch_phot}
\begin{figure}
	\centering
		\includegraphics[width=0.5\textwidth]{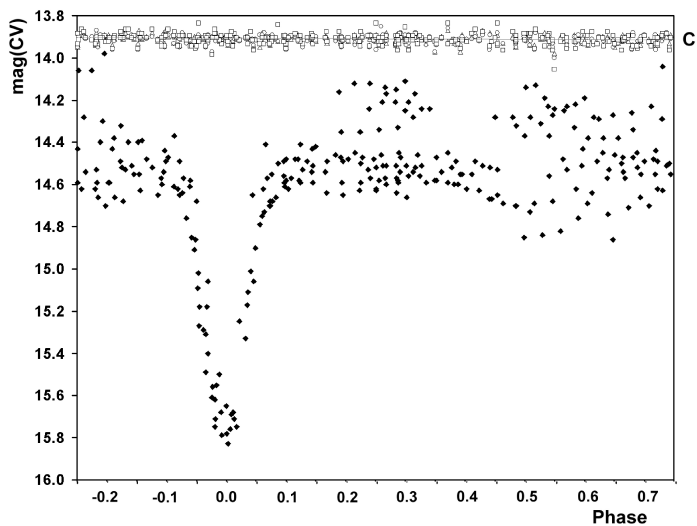}
	\caption{Phase plot of \astrobj{NSV 1907}, based on data from the Catalina Sky Survey \citep{Drak09}. The data have been folded with a period of $P = 0.2761060$ d \citep{Drak14}. Also included are the CSS light curves of three constant, nearby stars of similar magnitude (denoted as C); see text for details.}
	\label{Fig1}
\end{figure}

Archival photometry of \astrobj{NSV 1907} was procured from the CSS, which observed \astrobj{NSV 1907} during a timespan of about 2915 days. 283 observations of \astrobj{NSV 1907} are available in Data Release 2, which were downloaded from the corresponding website\footnote{http://nesssi.cacr.caltech.edu/DataRelease/}. Magnitudes derived from the CSS are unfiltered values that have been calibrated against $V$-band magnitudes and are designated hereafter as 'mag $(CV)$'. The CSS light curve is shown in Figure \ref{Fig0}. The corresponding phase plot, folded with the ephemeris given in \citet{Drak14}, is given in Fig. \ref{Fig1}. Note the obvious out-of-eclipse variability and the irregular depression suggestive of a shallow secondary minimum at phase $\varphi \approx 0.5$. Also included are the CSS light curves of three constant, nearby stars of similar magnitude (C1 = CSS J051653.6+033028; C2 = CSS J051648.0+033034; C3 = CSS J051651.3+033311; the light curves of C2 and C3 have been offset by respectively -1.0 mag and -1.8 mag to match the light curve of C1). The constancy of their light curves gives credit to the reality of the observed variability in the CSS light curve of \astrobj{NSV 1907}.

\subsection{New CCD Photometry} \label{new_phot}
To investigate the star in more detail, and to ensure that the observed secondary variability in CSS data is real and not an artefact, we have secured CCD photometric observations of \astrobj{NSV 1907} at three different sites. The instrumentations used are described in the following; an overview of the newly acquired photometry for \astrobj{NSV 1907} is given in Fig. \ref{Fig2}.

\begin{itemize}
	\item \textit{Private Observatory, Eckental, Germany}\\
	In 11 nights from 14/12/2014 to 15/12/2015, extended image series were acquired mostly under uneven sky conditions with a 250/10 SCT in a semi-automated mode and a SBIG ST8XME 
	camera. With 120s exposure time in the 2x2 binning mode, a total of 1369 measurements were taken (referred to hereafter as the Gr{\"o}bel dataset). To increase the S/N ratio, 
	no filter was used. Twilight sky-flat images were used for flatfield corrections. The reductions were performed with the MUNIWIN program package \citep{Motl11}.
	
	\begin{figure}[h!]
	\centering
		\includegraphics[width=0.5\textwidth]{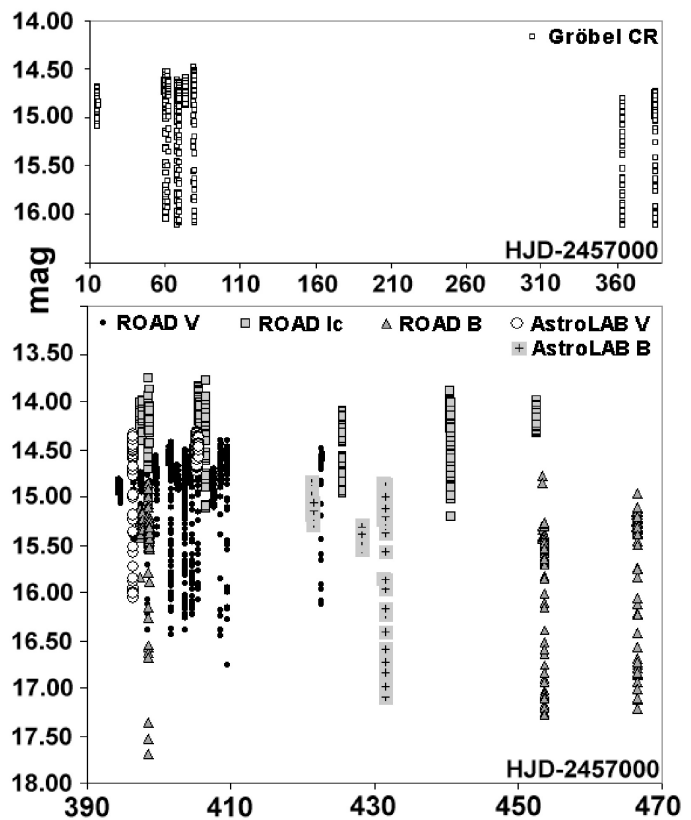}
	\caption{Light curve of \astrobj{NSV 1907}, based on our own photometric observations, as indicated. The (unfiltered) relative Gr{\"o}bel data have been shifted by $+14.8$ mag to match the $V$ observations.}
	\label{Fig2}
  \end{figure}

	\item \textit{Remote Observatory Atacama Desert (ROAD), Chile}\\
	Additional filtered observations were procured at the ROAD observatory \citep{Hamb12}, using an Orion Optics, UK Optimized Dall Kirkham 406/6.8 telescope and a FLI 16803 CCD 
	camera. Data were obtained in 20 nights from 07/01/2016 to 19/03/2016. Images were taken through Astrodon Photometric $B$, $V$, and $I\mathrm{C}$ filters. With 60s 
	exposure time in the 3x3 binning mode, a total of 2241 measurements were acquired (referred to hereafter as the ROAD dataset). Twilight sky-flat images were used for 
	flatfield corrections. The reductions were performed with the MAXIM DL program\footnote{http://www.cyanogen.com} and the determination of magnitudes using the LesvePhotometry 
	program\footnote{http://www.dppobservatory.net/}.
	\item \textit{AstroLAB, Ieper, Belgium}\\
	Further photometry was acquired at the AstroLAB observatory, Belgium, in 5 nights during the time span 08/01/2016 to 13/02/2016. The instrumental setup consisted of 
	an 684/4.4 Newton telescope on a Keller alt-azimuth mount and an SBIG STL-6303 E camera. Astrodon $B$ and $V$ filters were used. A total of 591 measurements were taken in the 3x3 binning mode, with either 60s (dataset 16/01/2016, $V$ filter; dataset 17/01/2016, $V$ filter) or 120s (dataset 08/01/2016, $V$ filter; dataset 02/02/2016, $B$ filter) 
	exposure time. The LesvePhotometry program was employed for the determination of magnitudes.
\end{itemize}

\subsection{Spectroscopic Observations} \label{spect_obs}
A spectrum of \astrobj{NSV 1907} was obtained on the night beginning on the 19th of January 2016 using the Intermediate Dispersion Spectrograph at the 2.5m Isaac Newton Telescope (INT) at the Roque de los Muchachos Observatory in La Palma, Canary Islands. The grating used was the R400V grating, a medium dispersion grating with a dispersion of 1.55 \AA \,per pixel. The central wavelength was 5513.6 \AA.

\begin{figure}[h!]
	\centering
		\includegraphics[width=0.47\textwidth]{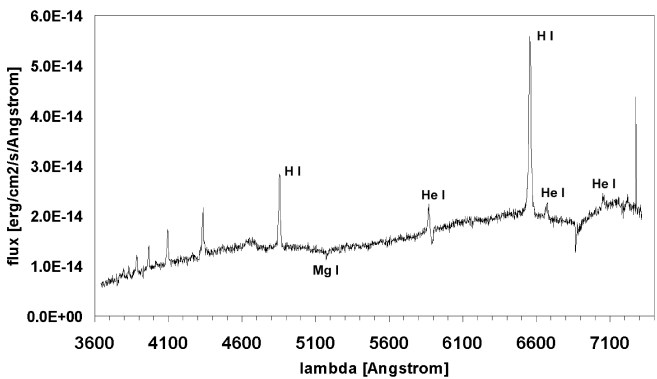}
	\caption{Spectrum of \astrobj{NSV 1907} obtained with the Isaac Newton Telescope (INT). See text for details.}
	\label{Fig3}
\end{figure}

The spectrum is shown in Figure \ref{Fig3}. It is the result of combining 3 individual exposures of 400s taken in succession. Since no suitable standard was available in that night's data, the spectrum was flux-calibrated using a standard, SP0946+139 (aka HD 84937), taken with the same instrument setup (grating, slit width and central wavelength) on the night of the 26th January 2016. The spectrum was reduced using the 'molly' reduction software\footnote{http://deneb.astro.warwick.ac.uk/phsaap/software/} written by Tom Marsh at the University of Warwick.

The spectrum (Fig. \ref{Fig3}) is typical for a cataclysmic variable, exhibiting the characteristic hydrogen and neutral helium emission lines. In addition to that, there is some evidence of higher excitation lines ('Bowen blend' at around λ4640 \AA, caused by ionised carbon and nitrogen, with possibly some contribution from He II emission at λ4686 \AA), which indicates the presence of a source of ionising photons (likely an accreting white dwarf). Except for the H$\alpha$ line, which shows some slight indication of asymmetry, the Balmer emission lines are symmetric and single-peaked, which is reminiscent of \astrobj{SW Sex} systems \citep{Thrs91} and argues against the presence of a high-inclination accretion disc. However, spectroscopic observations with higher resolution are needed to investigate this matter in detail.

The spectrum clearly exhibits secondary star features (Mg I 5168, 5174, 5185 \AA), which goes along well with the observed secondary eclipse (cf. section \ref{orbital_var}). This suggests a K spectral type for the secondary component, which would be expected for the observed long orbital period (\citealt{Pete05}; cf. also section \ref{period_study}). The Na D lines (5892, 5898 \AA) likely show an absorption component from the interstellar medium.

\subsection{Spectral Energy Distribution} \label{SED}
\begin{figure}[h!]
	\centering
		\includegraphics[width=0.47\textwidth]{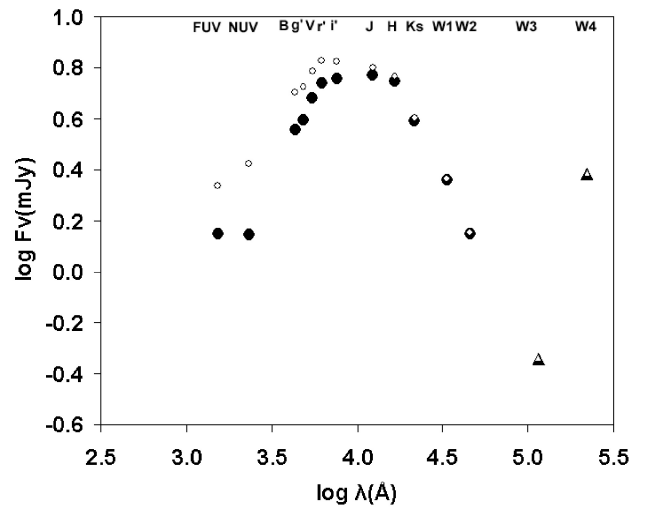}
	\caption{Spectral energy distribution of \astrobj{NSV 1907}, based on data obtained with the VizieR Photometry viewer. Filled symbols denote unreddened values, open symbols are based on extinction corrected values. Values in the WISE W3 and W4 bands denote upper limits and are represented by triangles. See text for details.}
	\label{FigSED}
\end{figure}

The spectral energy distribution of \astrobj{NSV 1907} is shown in Figure \ref{FigSED}, which has been based on data obtained with the VizieR Photometry viewer\footnote{http://vizier.u-strasbg.fr/vizier/sed/}. Both reddened (filled symbols) and unreddened (open symbols) values are shown. Values in the WISE W3 and W4 bands denote upper limits and are represented by triangles. Reddening estimates have been based on the calculations of \citet {Schl11} which should be considered as the maximum column density; extinction coefficients for the GALEX passbands were taken from \citet{Yuan13}. We estimate an interstellar extinction of $A\mathrm{V} \approx 0.27$ mag and $E(B-V) \approx 0.09$ mag in the line-of-sight to our target.

The shape of the SED is typical of nova-like variables, with the accreting WD contributing strongly to the short wavelength region. The secondary star contributes significantly at optical wavelengths, which is in agreement with its presumed spectral type and the observed secondary eclipse at an orbital phase of $\varphi \approx 0.5$.

\section{Data Analysis and Discussion} \label{data_analysis}

\subsection{Period Analysis} \label{period_study}
From our newly acquired observations and an analysis of CSS data, the following ephemeris could be derived, where phase zero corresponds to the primary eclipse.

\begin{equation} \label{eq1}
\resizebox{.83\hsize}{!}{$Min (HJD)$ = 2457060.3785(2) + 0.2761069(2) $\times E$}
\end{equation}

The corresponding times of minima are given in Table \ref{table_1}. The CSS light curve is not well sampled around the times of primary eclipse and the resulting errors in the times of minima are large; we have thus chosen to exclude CSS data from the period analysis.

\begin{table}[h!]
\caption{Observed times of minima.}
\label{table_1}
\begin{center}
\begin{tabular}{ccccc}
\hline
\hline 
Observer & HJD-2450000 & Error & Epoch & (O-C) \\ [0.5ex] 
 \hline\hline
 Gr{\"o}bel	& 7060.3790	& 0.00028 & 0	& 0.0005 \\ 
 \hline
 Gr{\"o}bel	& 7068.3850	& 0.00028 & 29 & -0.0006 \\ 
 \hline 
 Gr{\"o}bel	& 7070.3185 & 0.00022	& 36	& 0.0001 \\ 
 \hline
 Gr{\"o}bel	& 7080.2582 & 0.00014	& 72& 0.0000 \\ 
 \hline
 Gr{\"o}bel	& 7364.6485 & 0.00052	& 1102 & 0.0002 \\ 
 \hline
 Gr{\"o}bel	& 7387.5648 & 0.00036	& 1185 & -0.0004 \\ 
 \hline
 Dubois	& 7396.4006 & 0.00018	& 1217 & 0.0000 \\ 
 \hline
 Hambsch & 7401.6470 & 0.00034 & 1236	& 0.0004 \\ 
 \hline
 Hambsch & 7403.5791 & 0.00036 & 1243	& -0.0003 \\ 
 \hline
 Gr{\"o}bel	& 7425.3916	& 0.00018 & 1322	& -0.0002 \\ 
 \hline
 Hambsch & 7425.6677 & 0.00022 & 1323	& -0.0002\\
 \hline
 Dubois	& 7431.4666	& 0.00040 & 1344 & 0.0004 \\ [1ex]  
 \hline
\hline
\end{tabular}
\end{center}
\end{table}

\subsection{Additional Variability and Amplitudes} \label{orbital_var}
\begin{figure}[h!]
	\centering
		\includegraphics[width=0.47\textwidth]{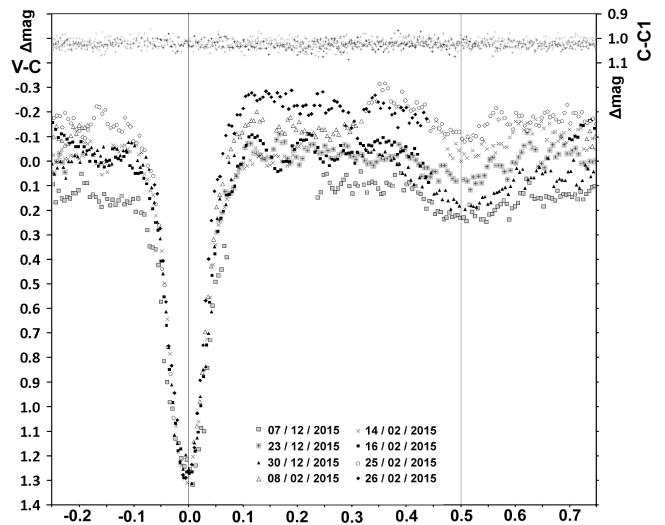}
	\caption{Phase plot of \astrobj{NSV 1907}, based on part of the unfiltered Gr{\"o}bel data (as indicated in the inset legend) and folded with the ephemeris given in Eq. (1). Note the obvious out-of-eclipse variability and the presence of a secondary minimum at phase $\varphi \approx 0.5$.}
	\label{Fig5}
\end{figure}

\begin{figure}[h!]
	\centering
		\includegraphics[width=0.47\textwidth]{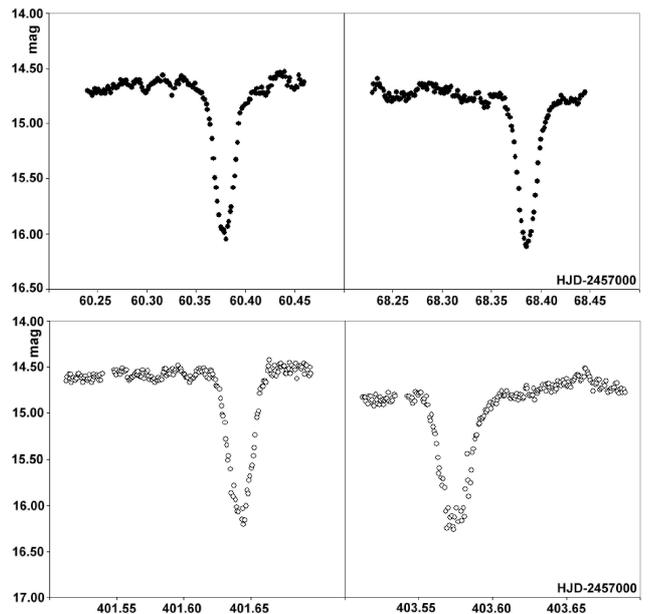}
	\caption{Detailed view of eclipses of \astrobj{NSV 1907}. The plots have been based on unfiltered Gr{\"o}bel data (upper panels; full circles) and ROAD $V$ data (lower panels; open circles). Note the short-term irregular variability that is present outside of eclipses.}
	\label{Fig6}
\end{figure}

From our new observations, the reality of the additional variability seen in CSS data in addition to the eclipses (short-term irregular variability, long-term variations) could be established, as well as the presence of a secondary minimum which seems to be a constant feature and takes the form of a small, irregular depression at phase $\varphi \approx 0.5$. The primary minimum seems to be slightly variable, too. Figure \ref{Fig5}, which has been based on part of the unfiltered Gr{\"o}bel data, nicely illustrates the observed additional variability. Detailed views of primary eclipses are presented in Figure \ref{Fig6}.

During our photometric coverage, the observed changes in mean brightness of the system outside of primary eclipse amounted to $\sim\,0.3$ mag ($CR$), $\sim\,0.5$ mag ($V$), and $\sim\,0.5$ mag ($B$), respectively. Archival data from the CSS indicate mean brightness variations of up to $0.7$ mag ($CV$), which becomes especially obvious during the active phase around HJD 2454500.

\begin{figure}[h]
	\centering
		\includegraphics[width=0.44\textwidth]{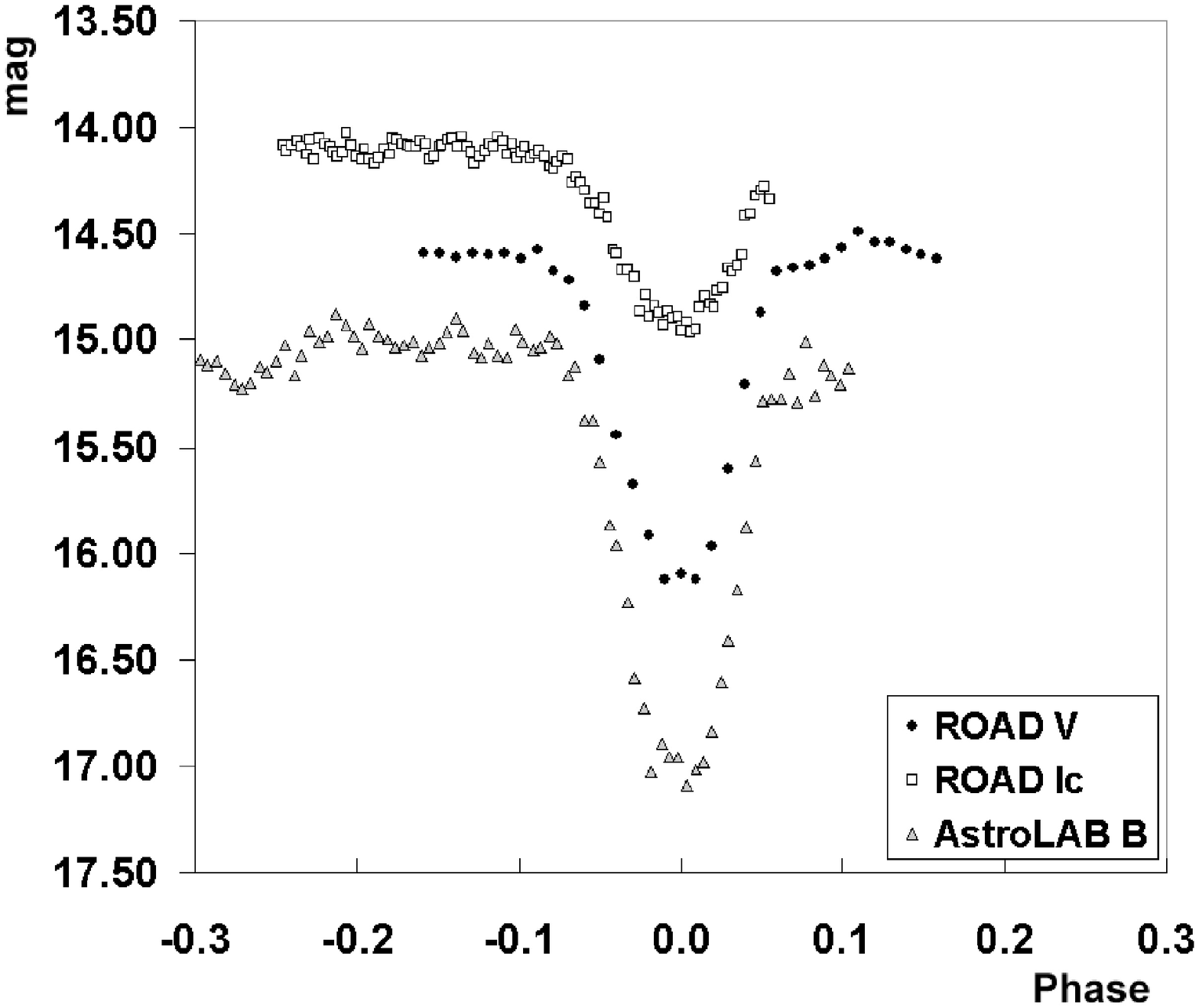}
	\caption{Detailed view of eclipses based on AstroLAB $B$ and ROAD $VI\mathrm{C}$ data. Outlying datapoints have been removed by visual inspection. The data were acquired on different orbital cycles (see text for details).}
	\label{Fig7}
\end{figure}

Unfortunately, no simultaneous multicolour coverage of the primary eclipse exists. However, judging from the long-term light curve, the primary eclipse is a rather stable feature. We have therefore derived approximate eclipse amplitudes employing the best-covered eclipses in $B$ (around HJD 2457431.46), $V$ (around 2457422.63), and $I\mathrm{C}$ (around HJD 2457425.67). An overview of these eclipses is presented in Fig. \ref{Fig7}. Although the scatter in the $B$ and $I\mathrm{C}$ observations is relatively large, it is obvious that the amplitude of the eclipse increases to shorter wavelengths ($\Delta B>\Delta V>\Delta I\mathrm{C}$), indicating that the hot object is eclipsed around phase $\varphi \approx 0$. We derive eclipse amplitudes of $\Delta B \approx 2.0$ mag, $\Delta V \approx 1.5$ mag, and $\Delta I\mathrm{C} \approx 0.8$ mag.

\begin{figure}[h!]
	\centering
		\includegraphics[width=0.45\textwidth]{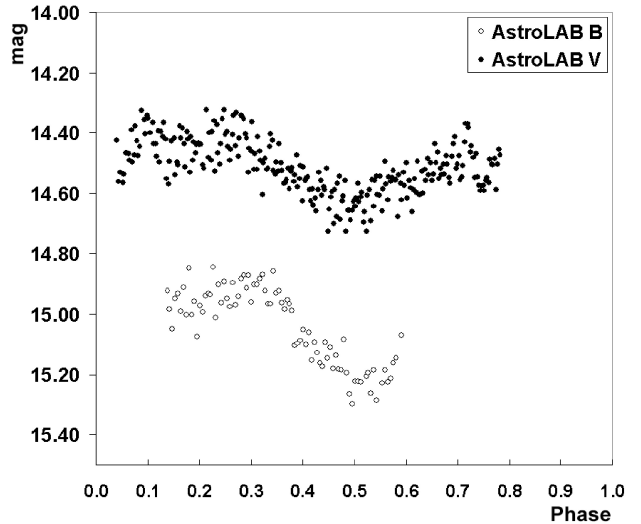}
	\caption{Detailed view of secondary eclipses based on AstroLAB $BV$ data. Outlying datapoints have been removed by visual inspection. The data were acquired on different orbital cycles (see text for details).}
	\label{Fig8}
\end{figure}

The presence of a secondary minimum in the light curve of \astrobj{NSV 1907} is remarkable as this is seldom observed in eclipsing cataclysmic variables. There are exceptions like e.g. the \astrobj{SW Sextantis} system \astrobj{V363 Aur} \citep{Thor04}. This star exhibits a secondary eclipse in the $R$ band light curve, which is absent in the $B$ band, implying that this feature is stronger at longer wavelengths. Apparently, the reverse holds true for \astrobj{NSV 1907}; from Fig. \ref{Fig8}, it becomes apparent that the amplitude of the secondary eclipse is greater in $B$ than in $V$.

Likely, then, different mechanisms are at work in \astrobj{NSV 1907}. It seems probable that the accretion disc occults the companion star that contributes a significant amount to the optical flux output of the system, which results in the observed secondary eclipse at phase $\varphi \approx 0.5$. The reflection effect (heating of the companion star by the WD primary and/or the accretion disc) may be important in this respect. There could also be material above and below the disc plane that contributes to the eclipse of the secondary.

\subsection{The 4.2 day period - the nodal precession period?} \label{nodal_precession_period}
\begin{figure}[h!]
	\centering
		\includegraphics[width=0.46\textwidth]{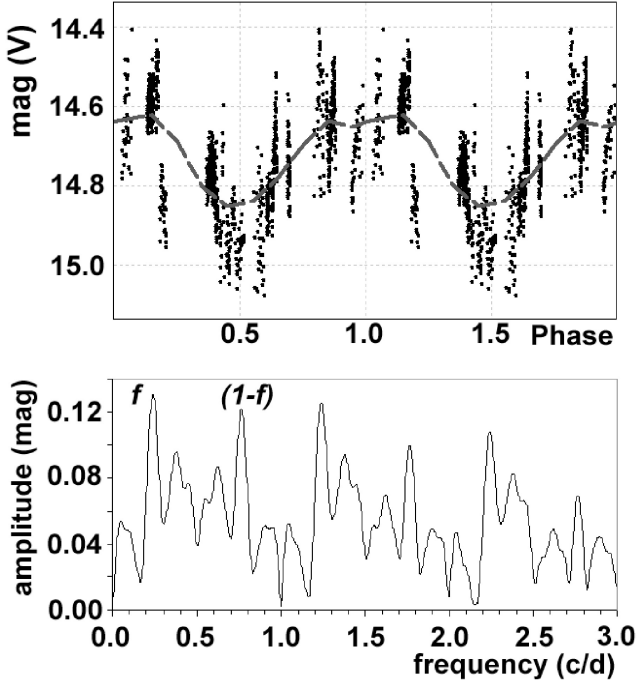}
	\caption{ROAD $V$ data, folded on the proposed 4.2 day ($f = 0.2375$ c/d) variations (upper panel; eclipses have been removed by visual inspection) and the corresponding power spectrum (lower panel), which is dominated by $f$, $(1-f$), and the corresponding daily aliases.}
	\label{Figps}
\end{figure}

We have searched for periodic signals other than the orbital period in the ROAD $V$ dataset, which covers the most consecutive orbital cycles. To this end, eclipses were removed by visual inspection and the resulting data were searched in the frequency range of $0 < f(c/d) < 3$ using \textsc{Period}04 \citep{Lenz05}. We find possible evidence for sinusoidal variations with a period of $P \approx 4.2$ d (Fig. \ref{Figps}), which is reminiscent of the findings of \citet{deMi16} who find a period of $P \approx 3.68$ d from an analysis of long-term photometric observations on the nova-like variable UX UMa. The authors interpret this signal as the signature of a retrograde precession of the accretion disc.

If we adopt the same interpretation for \astrobj{NSV 1907}, i.e. that the 4.2 day period is the nodal precession period, the superhump period $P\mathrm{sh}$ can be deduced from $f\mathrm{sh}$ = $f\mathrm{orb}$ + $f\mathrm{nodal}$. Assuming $P\mathrm{orb} = 0.2761$ d ($f\mathrm{orb} = 3.6219$ c/d) and $P\mathrm{nodal} = 4.211$ d ($f\mathrm{nodal} = 0.2375$ c/d), a superhump period of $P\mathrm{sh} = 0.2591$ d ($f\mathrm{sh} = 3.8594$ c/d) is derived. From this, the observed fractional period excess of the superhump period $\epsilon$ = ($P\mathrm{sh}$ - $P\mathrm{orb}$) / $P\mathrm{orb}$ \citep{Patt05} can be derived, which -- in the case of \astrobj{NSV 1907} -- amounts to $\epsilon \approx$ 0.06.

It has been shown that $\epsilon$ is strictly a function of the mass ratio $q$ = $M\mathrm{2}$/$M\mathrm{1}$ \citep{Patt05}. Using equation (8) of \citet{Patt05}, we derive $q \approx 0.25$. Considering the long period of \astrobj{NSV 1907}, which allows for a bigger and more massive secondary star than found in most other nova-like variables, this value seems reasonable \citep[cf. Fig 9 of][]{Patt05}.

However, because of the short time baseline of our data, these values have to be regarded as very preliminary only. More observations are needed to confirm the reality of the proposed period and verify its consistency over a longer timespan. Monitoring of the secondary eclipse might also be useful in establishing the presence of a tilted accretion disc in the system, in which case the secondary eclipse should be variable in time as a result of the changing projected area of the disc and WD system.

\section{Conclusion} \label{conclusion}
From our observations, we deduce that \astrobj{NSV 1907} is a deeply eclipsing, nova-like cataclysmic variable. Apart from the orbital variations (eclipses), changes in mean brightness and irregular short-term variability (flickering) were observed. The presence of a secondary minimum at phase $\varphi \approx 0.5$ was established. This is rarely observed in CVs and indicates that the contribution of the companion star is not entirely negligible as compared to the luminosity of the hot object. We have searched for periodic variability other than the orbital period and find possible evidence for sinusoidal variations with a period of $P \approx 4.2$ d, which we interpret as the nodal precession period. However, more observations are needed to confirm or disprove the reality of our finding.

No outbursts or VY Scl-like drops in brightness were observed either by the CSS or during our photometric monitoring. Because of its spectral characteristics (single-peaked emission lines, high-excitation lines) and the observed variability pattern (high-inclination, eclipsing system; V-shaped eclipses; mean brightness changes; flickering), we propose \astrobj{NSV 1907} as a new member of the \astrobj{SW Sextantis} class.

\begin{figure}[h!]
	\centering
		\includegraphics[width=0.47\textwidth]{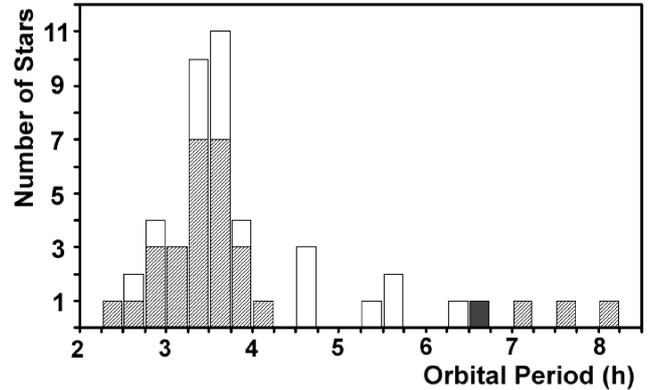}
	\caption{Orbital period distribution of \astrobj{SW Sex} stars, based on data from 'The Big List of SW Sextantis Stars' \citep{Hoar03}. Only definite \astrobj{SW Sex} stars (dashed area) and probable candidates (white area) have been considered. The position of \astrobj{NSV 1907} is indicated by the dark area.}
	\label{Fig_big_list}
\end{figure}

In fact, the star's properties and light curve are strongly reminiscent of other long-period \astrobj{SW Sex} systems like e.g. \astrobj{V363 Aur} (the longest-period definite \astrobj{SW Sex} star with $P\mathrm{orb} \approx 7.71$ h), which, incidentally, also exhibits a conspicuous secondary eclipse and secondary star features in its spectrum \citep{Thor04}. Figure \ref{Fig_big_list} shows the orbital period distribution of \astrobj{SW Sex} stars. The plot has been based on data from the 'Big List of SW Sextantis Stars' \citep{Hoar03}. Only definite members of the group and probable candidates have been considered. With an orbital period of $P \approx 6.63$ hours, \astrobj{NSV 1907} -- if confirmed as a member -- would be the \astrobj{SW Sex} star with the fourth longest period.

Further photometric and spectroscopic observations are encouraged. Time-resolved spectroscopy, in particular, should prove valuable in investigating class-defining features such as the radial velocity variations of the emission lines, which exhibit maximum blueshift near an orbital phase of $\varphi\mathrm{orb} \approx 0.5$, or the typical displacement of the radial velocity curve of the line wings \citep{Rodr07a}.

\section*{Acknowledgements}
\small
We thank Tom Marsh at the University of Warwick for the use of the 'molly' reduction software and Marek Skarka for helpful discussions. The research leading to these results has received funding from the European Research Council under the European Union's Seventh Framework Programme (FP/2007-2013) / ERC Grant Agreement n. 320964 (WDTracer). This research has made use of data from the Catalina Sky Survey. The CSS survey is funded by the National Aeronautics and Space Administration under Grant No. NNG05GF22G issued through the Science Mission Directorate Near-Earth Objects Observations Program. The CRTS survey is supported by the U.S.~National Science Foundation under grants AST-0909182. Furthermore, use was made of the SIMBAD database and the VizieR catalogue access tool operated at CDS, Strasbourg, France.


\begin{thebibliography}{0}
\expandafter\ifx\csname natexlab\endcsname\relax\def\natexlab#1{#1}\fi

\end{thebibliography}


\section*{References}

\begin{thebibliography}{}

\bibitem[\protect\citeauthoryear{Cusumano et al.}{2010}]{Cusu10} Cusumano, G., La Parola, V., Segreto, A. et al. 2010, A\&A, 524, 64
\bibitem[\protect\citeauthoryear{de Miguel et al.}{2016}]{deMi16} de Miguel, E., Patterson, J., Cejudo, D. et al. 2016, \mnras, 457, 1447
\bibitem[\protect\citeauthoryear{Drake et al.}{2009}]{Drak09} Drake, A. J., Djorgovski, S. G., Mahabal, A. et al. 2009, \apj, 696, 870
\bibitem[\protect\citeauthoryear{Drake et al.}{2014}]{Drak14} Drake, A. J., Graham, M. J., Djorgovski, S. G. et al. 2014, \apjs, 213, 9
\bibitem[\protect\citeauthoryear{Hambsch}{2012}]{Hamb12} Hambsch, F.-J. 2012, JAVSO, 40, 1003
\bibitem[\protect\citeauthoryear{Hellier}{2001}]{Hell01} Hellier, C. 2001, Cataclysmic Variable Stars, Springer, London
\bibitem[\protect\citeauthoryear{Hoard et al.}{2003}]{Hoar03} Hoard, D. W., Szkody, P., Froning, C. S. et al. 2003, \aj, 126, 2473
\bibitem[\protect\citeauthoryear{Hoffmeister}{1963}]{Hoff63} Hoffmeister, C. 1963, Astronomische Nachrichten, 287, 169
\bibitem[\protect\citeauthoryear{Kukarkin \& Kholopov}{1982}]{KuKh82} Kukarkin, B. V., Kholopov, P. N. 1982, Publication Office Nauka, Moscow, 287, 169
\bibitem[\protect\citeauthoryear{Lenz \& Breger}{2005}]{Lenz05} Lenz, P., \& Breger, M. 2005, Communications in Asteroseismology, 146, 53
\bibitem[\protect\citeauthoryear{Martin et al.}{2005}]{Mart05} Martin, D. C., Fanson, J., Schiminovich, D. 2005, \apj, 619, 1
\bibitem[\protect\citeauthoryear{Morrissey et al.}{2007}]{Morr07} Morrissey, P., Conrow, T., Barlow, T.~A. et al.\ 2007, \apjs, 173, 682 
\bibitem[\protect\citeauthoryear{Motl}{2011}]{Motl11} Motl, D. 2011, C-Munipack, http://c-munipack.sourceforge.net
\bibitem[\protect\citeauthoryear{Patterson et al.}{2005}]{Patt05} Patterson, J., Kemp, J., Harvey, D. A. 2005, \pasp, 117, 1204
\bibitem[\protect\citeauthoryear{Peters \& Thorstensen}{2005}]{Pete05} Peters, C.~S., \& Thorstensen, J.~R.\ 2005, \pasp, 117, 1386
\bibitem[\protect\citeauthoryear{Rodr{\'{\i}}guez-Gil et al.}{2007a}]{Rodr07a} Rodr{\'{\i}}guez-Gil, P., Schmidtobreick, L., G{\"a}nsicke, B.~T.\ 2007, \mnras, 374, 1359
\bibitem[\protect\citeauthoryear{Rodr{\'{\i}}guez-Gil et al.}{2007b}]{Rodr07b} Rodr{\'{\i}}guez-Gil, P., G{\"a}nsicke, B.~T., Hagen, H.-J. et al.\ 2007, \mnras, 377, 1747
\bibitem[\protect\citeauthoryear{Schlafly \& Finkbeiner}{2011}]{Schl11} Schlafly, E.~F., \& Finkbeiner, D.~P.\ 2011, \apj, 737, 103
\bibitem[\protect\citeauthoryear{Schmidtobreick et al.}{2012}]{Schm12} Schmidtobreick, L., Rodr{\'{\i}}guez-Gil, P., G{\"a}nsicke, B.~T. 2012, MmSAI, 83, 610
\bibitem[\protect\citeauthoryear{Shappee et al.}{2014}]{Shap14} Shappee, B. J., Prieto, J. L., Grupe, D. 2014, \apj, 788, 48
\bibitem[\protect\citeauthoryear{Thoroughgood et al.}{2004}]{Thor04} Thoroughgood, T. D., Dhillon, V. S., Watson, C. A. 2004, \mnras, 353, 1135
\bibitem[\protect\citeauthoryear{Thorstensen et al.}{1991}]{Thrs91} Thorstensen, J. R., Davis, M. K., Ringwald, F. A. 1991, \aj, 102, 683
\bibitem[\protect\citeauthoryear{Ubertini et al.}{2003}]{Uber03} Ubertini, P., Lebrun, F., Di Cocco, G. et al.\ 2003, \aap, 411, L131 
\bibitem[\protect\citeauthoryear{Voges et al.}{1991}]{Voge91} Voges, W., Aschenbach, B., Boller, Th. 1999, A\&A, 349, 389
\bibitem[\protect\citeauthoryear{Warner}{1995}]{Warn95} Warner, B. 1995, Cataclysmic Variable Stars, Camb. Astrophys. Ser., Vol. 28
\bibitem[\protect\citeauthoryear{Yuan et al.}{2013}]{Yuan13} Yuan, H.~B., Liu, X.~W., \& Xiang, M.~S.\ 2013, \mnras, 430, 2188
	
\end{thebibliography}
\end{document}